\documentstyle{sup2}
\include{psfig}
\vspace{-8pt}

\title[Superlattices and Microstructures, Vol.\ ??, No.\ ?, 1999]
{Resonant transport through midgap states in voltage-biased Josephson
junctions of $d$-wave superconductors
}
\author[Superlattices and Microstructures, Vol.\ ??, No.\ ?, 1999]
{Tomas~L\"ofwander, G\"oran~Johansson,\cr\vspace{10pt}
Vitaly~Shumeiko, G\"oran~Wendin, and Magnus Hurd
\cite{hurd_address}\cr\vspace{10pt}
{\normalsize\it Chalmers University of Technology and G\"oteborg University,
S-412 96 G\"oteborg, Sweden}\cr
}

\begin{document}
\label{firstpage}
\maketitle
\sloppy
\begin{center}
\received{(Received ????)}
\end{center}
%************************************************************************%

\begin{abstract}
  We study theoretically the ac Josephson effect in voltage biased
  planar junctions of $d$-wave superconductors. For some orientations
  of the superconductors a current peak is found at finite voltage in
  the current-voltage characteristics. We pick out the relevant
  physical processes and write down an analytical formula for the
  current which clearly shows how the midgap state acts as a resonance
  and produces the peak. We present a possible explanation for the
  zero-bias conductance peak, recently found in experiments on grain
  boundary junctions of high-temperature superconductors, in terms of
  resonant transmission through midgap state of quasiparticles
  undergoing multiple Andreev reflections. We note that within our
  framework the zero-bias conductance peak appears in rather
  transparent Josephson junctions of $d$-wave superconductors.
\end{abstract}

\section{Introduction}
The controversy regarding the symmetry of the order parameter in
high-temperature superconductors (HTS) seems today, to some extent,
have been resolved in favour of the $d_{x^2-y^2}$-wave ($d$-wave)
symmetry~\cite{Harl,Kouz,Tsue3}. Since current transport through
superconducting junctions is sensitive to both magnitude and phase of
the order parameter, experiments on Josephson junctions can measure
the intrinsic phase of the $d$-wave order parameter. Therefore, a
great deal of attention has been directed towards the Josephson effect
in HTS junctions. Both the dc~\cite{BBR,TanKas3,TanKas4,RiedBag} and
ac~\cite{Hurd,LJHW,HLJW,BarSvi,SamDat} effects have been studied.

An important discovery was that at surfaces and interfaces of $d$-wave
superconductors a midgap state (MGS) may be formed~\cite{Hu}, which
affects the current transport. A quasiparticle reflected at the
interface will sense different order parameters before and after the
scattering event since the momentum is changed. When there is a sign
difference between the order parameters before and after scattering, a
zero-energy bound state is formed (MGS). In normal metal-$d$-wave
superconductor (${\mbox N}|d$) junctions the MGS gives rise to a
zero-bias conductance peak (ZBCP), established both
theoretically~\cite{Hu,TanKas1,FRS} and
experimentally~\cite{GXL,Cov,SinNg,Alff1}. Previously, it has been
shown that MGS gives rise to a current peak at finite voltage in both
$s$-wave superconductor-$d$-wave superconductor ($s|d$) and $d$-wave
superconductor-$d$-wave superconductor ($d|d$) junctions, resulting in
negative differential conductance~\cite{Hurd,HLJW,BarSvi,SamDat}. The
main purpose of this paper is to analyze the results of
Refs.~\cite{Hurd} and \cite{HLJW} in terms of the physical processes
giving rise to the peak at finite voltage.

Knowing the effects of MGS in the $N|d$ junction, it is not
straightforward to predict the effects in the $s|d$ and $d|d$
junctions. This is the case because the mechanisms involved in the
current transport are more complicated when both electrodes are
superconducting. In a voltage biased junction between two
superconductors, the phase difference over the junction is
time-dependent according to the Josephson relation
$\dot{\varphi}=2eV/\hbar$. This makes the scattering in the junction
inelastic: a quasiparticle incident on the junction at energy $E$
produces a scattering state with sideband energies $E_n=E+neV$, where
$V$ is the voltage over the junction and $n$ is an integer.

The knowledge about current transport through voltage biased $s$-wave
$s|s$ junctions has greatly increased in recent years. Using a
Landauer-B\"uttiker scattering method (extended to include
superconducting electrodes) it has become possible to understand $s|s$
junctions with arbitrary transparency of the insulator in terms of
multiple Andreev reflections~\cite{BSW,AveBar,SBW,BSBW,JWBS}. The
theory has also been extended to include $d$-wave
superconductors~\cite{Hurd,HLJW}. This theory is used in the present
paper in order to study the effects of MGS in voltage biased
superconducting junctions.

The rest of the paper is organized in the following way. In Chapter 2
we briefly describe the model and how to calculate the current-voltage
(IV) characteristics. In Chapter 3, the scattering problem is solved
analytically taking into account only processes giving rise to the
current peak. We study the $s|d$ junction (where the peak is most
pronounced) and compare the current contribution from the processes
responsible for the current peak with the complete (numerically
calculated) IV-curve. In Chapter 4 we present a qualitative discussion
of how a ZBCP may appear in voltage biased $d|d$ junctions. This study
was motivated by recent experiments~\cite{Alff2,Alff3}. In Chapter 5
we summarize the paper.

\section{The model}

\begin{figure}[t]
  \centerline{\psfig{figure=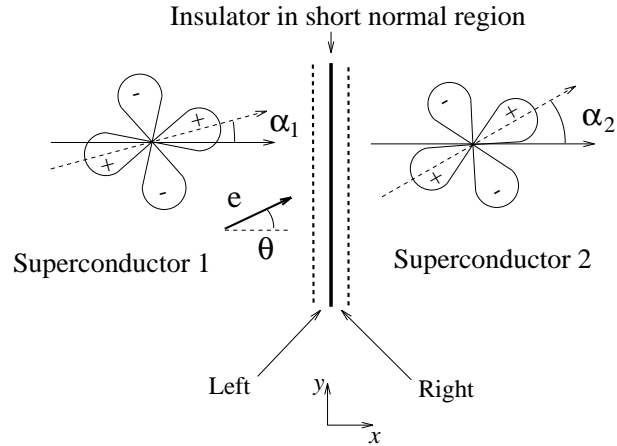,width=8cm}}
  \caption{The model system we are considering. The orientations of
    the two $d$-wave superconductors are given by the angles
    $\alpha_1$ and $\alpha_2$. An electron-like quasiparticle is
    incident on the junction at the angle $\theta$. We have here
    introduced short normal regions on each side of the barrier. The
    strength of the barrier can be tuned from the ballistic to the
    insulating case.}
  \label{system_fig}
\end{figure}

The system we are considering is a planar junction between $d$-wave
superconductors as shown in Fig.~\ref{system_fig}. The superconducting
gap function depends on the quasiparticle's direction of propagation
$\theta$. We assume that the superconductors have pure $d$-wave
symmetry and let $\Delta(\theta)=\Delta_0\cos[2(\theta-\alpha)]$,
where $\alpha$ is the orientation of the superconductor relative to
the interface normal.  For an $s$-wave superconductor,
$\Delta(\theta)=\Delta_s$.

We neglect surface roughness in our treatment of the junction region,
modelling the barrier by the potential $V(x)=H\delta(x)$. The
reflection and transmission amplitudes of the barrier are therefore
angle-dependent: $r(\theta)=Z/(i\cos\theta-Z)$ and
$t(\theta)=i\cos\theta/(i\cos\theta-Z)$, where the dimensionless
parameter $Z=2mH/\hbar^2$ describes the transparency of the
junction~\cite{Brud}. A small value of $Z$ corresponds to a
transparent junction and a large value corresponds to a tunnel
junction. The limiting cases are $Z=0$ (the ballistic junction - no
scattering) and $Z=\infty$ (two uncoupled electrodes). Having only
specular reflections implies that the quasiparticle's momentum along
the junction $k_y$ is a conserved quantity. This means that the
current is an average over all injection angles where the current
contribution from each injection angle can be calculated separately.
The difference from a junction between $s$-wave superconductors is
that we here must take into account the angle dependence of the
$d$-wave gap. In particular, we must remember that the gap changes
after normal reflection at the barrier.

Self-consistency of the gap is important in junctions of
$d$-superconductors~\cite{BarSvi}. It has been shown in the tunnelling
limit that including self-consistency of the gap produces bound states
with non-zero energy along with the MGS.  The length of the normal
regions introduced on each side of the barrier in
Fig.~\ref{system_fig} could be thought of as a model of the possible
suppression of the $d$-wave gap near the interface. Since we in this
paper focus on effects of MGS, which is due to the intrinsic sign of
the $d$-wave order parameter only, we neglect these effects and put
the length of the normal regions to zero. We therefore assume a
step-like behaviour of the gap-functions:
\begin{equation}
\Delta=\left\{
\begin{array}{ll}
\Delta_1(\theta), & x<0\nonumber\\
\Delta_2(\theta)e^{i \phi_0}, & x>0.
\end{array}
\right.
\label{order_parameter}
\end{equation}
Since the overall phase is unimportant we can here choose $\Delta_1$
real and let the phase of the right-hand superconductor be equal to
the phase-difference $\phi_0$ over the junction.

The time-dependent Bogoliubov-de Gennes equation describing
anisotropic superconductors is solved for a voltage biased junction,
modeled as described above, by the method described in detail in
Ref.~\cite{HLJW}. We obtain the scattering states by matching ansatz
wavefunctions at the NS interfaces and at the barrier for all
injection angles $\theta$ and energies $E$. The current is a sum over
contributions from all scattering states, where each contribution is
found by inserting the wavefunction into the current formula. The sum
is in the end turned into an integration over injection angle and
energy.

We choose to calculate the current in the normal region to the left of
the barrier. The wave function in this region is
\begin{equation}
{\Psi}_{L}^{\sigma} =
\sum_{n} 
\left(
\begin{array}{c}
a_n^{\sigma} e^{i {\bf k}^e\cdot{\bf x}}
+d_n^{\sigma} e^{i {\bf\bar{k}}^e\cdot{\bf x}}\\
b_n^{\sigma} e^{i {\bf k}^h\cdot{\bf x}} 
+c_n^{\sigma} e^{i {\bf \bar{k}}^h\cdot{\bf x}} 
\end{array}
\right)
e^{-i(\frac{E_n t}{\hbar}+\frac{n\phi_0}{2})},
\label{Psi_L}
\end{equation}
where the side-band energy is $E_n=E+neV$. The index
$\sigma=\left\{e^{\rightarrow}, h^{\rightarrow}, h^{\leftarrow},
  e^{\leftarrow}\right\}$ labels the four types of incoming
quasiparticles: electron-like and hole-like quasiparticles injected
from the left and right electrodes. Here is ${\bf k}^{e/h}$ the wave
vector of electrons and holes respectively. In our two-dimensional
problem we have ${\bf k}=(k_x,k_y)=k(\cos\theta,\sin\theta)$ and
${\bar{\bf k}}=(-k_x,k_y)=k(\cos{\bar \theta},\sin{\bar \theta})$,
where ${\bar \theta}=\pi-\theta$ is the angle after normal reflection.
The wave function coefficients $a$, $b$, $c$, and $d$ in
Eq.~(\ref{Psi_L}) are found by solving the matching equations.

The current per $ab$-plane for a junction of width $L_y$ is
\begin{equation}
\frac{I}{\sigma_0}=\frac{1}{4D}\int_{-\pi/2}^{\pi/2}d\theta\cos\theta
\int_{-\infty}^{\infty} dE f(E)
\sum_{\sigma}j^{\sigma}(\theta,E),
\label{current}
\end{equation}
where $\sigma_0=2ek_F L_y D/\pi h$ is the normal state conductance,
$D=\int d\theta {\cal D}\cos\theta/2$ is the normal state transparency
averaged over all angles $\theta$ (${\cal D}=|t(\theta)|^2$), $E_F$ is
the Fermi energy, and $f(E)=1/[1+\exp(E/k_BT)]$ is the Fermi
distribution function. In Eq.~(\ref{current}) we sum over
electron-like and hole-like quasiparticles injected from the left and
right reservoirs described by the index $\sigma$.

The current density appearing in Eq.~(\ref{current}) is expressed in
terms of the wavefunction coefficients in $\Psi_L^{\sigma}$:
\begin{equation}
j^{\sigma}(\theta,E)=N^{\sigma}(\theta,E)\sum_n
|a_n^{\sigma}|^2-|d_n^{\sigma}|^2
+|b_n^{\sigma}|^2-|c_n^{\sigma}|^2.
\label{current_density}
\end{equation}
In order to collect contributions with the same momentum along the
junction $k_y$, we inject the four quasiparticles $\sigma$ at four
different angles with the following four gaps:
$\Delta^{\sigma}=\left\{\Delta_1, {\bar \Delta}_1, \Delta_2, {\bar
    \Delta}_2\right\}$ suppressing the dependence on $\theta$. A
consequence of this is that four density of states
$N^{\sigma}(\theta,E)=\Theta(|E|-|\Delta^{\sigma}(\theta)|)|E|/
\sqrt{E^2-\Delta^{\sigma}(\theta)^2}$ appear in
Eq.~(\ref{current_density}).

\section{Midgap state resonance}

Consider an ${\mbox N}|d_{\pi/4}$ junction biased at arbitrary small
voltage $V$ and an electron incident on the junction from the normal
metal side at the Fermi energy. Due to MGS, the electron will be
Andreev reflected with unit probability, independent of the strength
of the interface barrier. In energy space, this can be thought of as
resonant {\it transmission} through MGS: the electron at energy $E$ is
turned into a hole with energy $E+2eV$ with unit probability. In the
Andreev reflection process a charge $2e$ is transferred into the
superconductor (forming a Cooper pair) and a current flows through the
junction. This current flow through MGS is the explanation of the ZBCP
seen in many experiments~\cite{GXL,Cov,SinNg,Alff1}.

The advantage of the above description of current flow through MGS is
that it can be generalized to the case when the normal metal is
changed to a superconductor. In a junction between two superconductors
an incident quasiparticle will create both electrons and holes in the
normal region, which will undergo multiple Andreev reflections.
Because of the voltage drop over the normal region each particle
changes its energy by $eV$ each time it passes the normal region. In
this way a scattering state with amplitudes at all sideband energies
$E_n$ is built up~\cite{BSW}. If we consider a quasiparticle incident
from the left superconductor at energy $E$, Andreev reflections at the
right superconductor will take place at the odd sideband energies
$E_{2n+1}$, while Andreev reflections at the left superconductor will
take place at the even sideband energies $E_{2n}$.  When e.g. the
right electrode is a $d$-wave superconductor, oriented in such a way
that MGS is formed, a {\it transmission} resonance is produced in
energy space: the probability of transmission from energy $E_{2n}$ to
energy $E_{2n+2}$ is unity if $E_{2n+1}=0$.

The resonant transport through MGS described above results in current
peaks in the IV-characteristics of junctions containing $d$-wave
superconductors, as previously reported in Refs.~\cite{Hurd} and
\cite{HLJW}. Since the current peak is most pronounced in the
$\mbox{s}|\mbox{d}_{\pi/4}$ junction we will here concentrate on this
particular junction.

\subsection{Midgap state resonance in the $s|d_{\pi/4}$ junction}

\begin{figure}[t]
\centerline{\psfig{figure=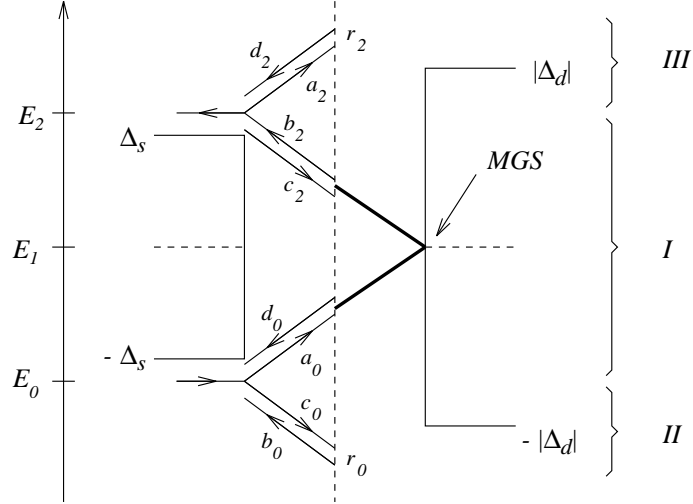,width=9cm}}
\caption{The two-particle process in the $s|d_{\pi/4}$ junction
  involving MGS for a quasiparticle injected from the left electrode.
  When matching, we divide the scattering state into three parts: the
  part in between the injection point and the exit point (region I),
  the part below the injection point (region II), and the part above
  the exit point (region III). The MGS resonance is marked by the
  thick lines.}
\label{MGSres_fig}
\end{figure}

Generally, it is possible to express the dc current in superconducting
junctions as a sum over contributions from different $n$-particle
processes, where $n$ is an integer. In a single scattering state,
originating from a quasiparticle incoming at energy $E$, the
$n$-particle contribution to the dc current is $n \cdot I^{p}_{n}$,
where $ I^{p}_{n}$ is the outgoing probability current at energy
$E_{n}=E+neV$ \cite{JWBS}. In this section we will focus on the
resonant two-particle process involving the MGS and derive the
expression for its contribution to the current explicitly, without
using the concept of probability current. To be specific we will
calculate the dc current contribution from quasiparticles incoming
from the left, at energy $E \approx -eV$, that are Andreev reflected
through the MGS on the right hand side, at $E_1\approx 0$ and then
leave the normal region to the left at energy $E_2\approx eV$ (see
Fig.~\ref{MGSres_fig}). This two-particle process involves one Andreev
reflection, implying a net current transport of $2e$, as we will also
see in the final expression for the current.

\begin{figure}[t]
\centerline{\psfig{figure=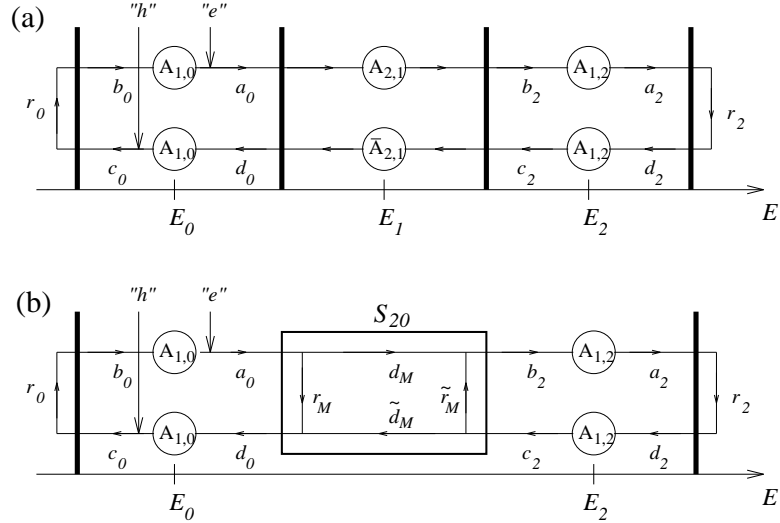,width=10cm}}
\caption{Mapping of the process in Fig.~\ref{MGSres_fig} on transport
  through a one-dimensional potential barrier structure in
  energy-space. In (a) the complete structure of the scattering state
  is shown, collecting the reflections from outside regions I-III into
  the reflection amplitudes $r_0$ and $r_2$. In (b) the transport
  through MGS has been collected into a scattering matrix.}
\label{sd_map_fig}
\end{figure}

In order to solve the scattering problem analytically, we divide the
energy axis into three parts: the part in between the injection point
and the exit point, the part below the injection point, and finally
the part above the exit point. We denote these three parts by I, II,
and III, as shown in Fig.~\ref{MGSres_fig}. The problem of calculating
the scattering state can be mapped onto the problem of calculating the
wavefunction for tunneling through a one-dimensional multi-barrier
structure (in energy space), as drawn in Fig.~\ref{sd_map_fig}. The
structure of the scattering state is better illustrated by
Fig.~\ref{sd_map_fig}, which also makes clear the way the matching of
the ansatz wave functions is done and how resonances may appear. From
now on, the discussion will therefore be connected to
Fig.~\ref{sd_map_fig}.

The coefficients in part I are coupled by a scattering matrix
$S_{20}$:
\begin{equation}
\left(
\begin{array}{c}
d_0^{\sigma}\\
b_2^{\sigma}
\end{array}
\right)
=S_{20}
\left(
\begin{array}{c}
a_0^{\sigma}\\
c_2^{\sigma}
\end{array}
\right)
=\left(
\begin{array}{cc}
r_M & {\tilde d}_M\\
d_M & {\tilde r}_M
\end{array}
\right)
\left(
\begin{array}{c}
a_0^{\sigma}\\
c_2^{\sigma}
\end{array}
\right),
\label{s_matrix}
\end{equation}
where the reflection amplitude ${\tilde r}_M$ is related to the other
amplitudes by ${\tilde r}_M=-r_M^*{\tilde d}_M/d_M^*$. By matching
inside region I, we find the amplitudes for reflection and
transmission through MGS: $d_M=|t|^2 A_{2,1}/(1-A_{2,1}{\bar
  A}_{2,1}|r|^2)$, ${\tilde d}_M=|t|^2{\bar A}_{2,1}/(1-A_{2,1}{\bar
  A}_{2,1}|r|^2)$, $r_M=r(1-A_{2,1}{\bar A}_{2,1})/(1-A_{2,1}{\bar
  A}_{2,1}|r|^2)$, and ${\tilde r}_M=r^*(1-A_{2,1}{\bar
  A}_{2,1})/(1-A_{2,1}{\bar A}_{2,1}|r|^2)$. Here we have introduced
the amplitude $A_{i,n}=A_i(E_n,\theta)$ for Andreev reflection at
energy $E_n$, which is defined in terms of the BCS coherence factors
$u$ and $v$ as
\begin{eqnarray}
A_i(E,\theta)=\frac{v_i(E,\theta)}{u_i(E,\theta)}=
\frac{E-sgn(E)\sqrt{E^2-\Delta_i(\theta)^2}}{\Delta_i(\theta)},
\hspace{1cm}|E|>|\Delta_i(\theta)|,\nonumber\\
A_i(E,\theta)=\frac{v_i(E,\theta)}{u_i(E,\theta)}=
\frac{E-i\sqrt{\Delta_i(\theta)^2-E^2}}{\Delta_i(\theta)},
\hspace{1cm}|E|<|\Delta_i(\theta)|,
\end{eqnarray}
where $|u_i|^2+|v_i|^2=1$ and $i=1$, $2$ refers to the left and right
superconductors respectively. The barred amplitudes ${\bar A}$ are
calculated at the angle ${\bar\theta}=\pi-\theta$.

The amplitudes for reflection from energies outside regions I, II, and
III, all the way to $\pm\infty$ in energy (taking into account all
possible processes) can be collected into $r_0$ and $r_2$
respectively. The exact numerical values of $r_0$ and $r_2$ are
determined by solving the matching equations for the complete
scattering state by the method described in Ref.~\cite{HLJW}. Without
solving these rather complex equations analytically, we can write down
the following formal relations between coefficients within part II and
III respectively:
\begin{equation}
\begin{array}{cc}
b_0^{\sigma}=r_0 c_0^{\sigma}, & d_2^{\sigma}=r_2 a_2^{\sigma}.
\end{array}
\end{equation}

At the injection and exit points, we get four equations connecting
part I with the parts II and III:
\begin{equation}
\begin{array}{cc}
a_0^{\sigma}=J\delta_{\sigma e^{\rightarrow}}+A_{1,0}b_0^{\sigma},
& c_0^{\sigma}=J\delta_{\sigma h^{\rightarrow}}+A_{1,0}d_0^{\sigma}\\
a_2^{\sigma}=A_{1,2}b_2^{\sigma},
& c_2^{\sigma}=A_{1,2}d_2^{\sigma},\label{abcd}
\end{array}
\end{equation}
where $J=[u_1(E)^2-v_1(E)^2]/u_1(E)$ is the amplitude for injection
into the normal region from the left superconductor. The
$\delta$-functions in Eq.(\ref{abcd}) are included because we do the
matching for incoming electron-like ($\sigma=e^{\rightarrow}$
$\Rightarrow$ $\delta_{\sigma e^{\rightarrow}}=1$) and hole-like
($\sigma=h^{\rightarrow}$ $\Rightarrow$ $\delta_{\sigma
  h^{\rightarrow}}=1$) quasiparticles separately. The two
contributions are summed up in the end in the current formula.  The
contributions to the two-particle current from particles injected from
the right superconductor will be proportional to ${\cal D}^2$ since
the trajectories are non-resonant (they do not hit MGS since it only
appears at the surface of the right superconductor). We therefore
neglect these contributions to the current in the present discussion.
The structure of the above matching equations is drawn in
Fig.~\ref{sd_map_fig}b.

Solving Eqs.~(\ref{s_matrix})-(\ref{abcd}), we get the following
expression for the current density by inserting the obtained
coefficients $a_n$, $b_n$, $c_n$, and $d_n$ ($n=0$, $2$) into
Eq.~(\ref{current_density}):
\begin{equation}
\sum_{\sigma} j^\sigma(\theta,E)\approx
\sum_{\sigma=e^{\rightarrow}, h^{\rightarrow}} j^{\sigma}(\theta,E)=
2(1-A_{1,0}^2)(1-A_{1,2}^2)|G_{20}|^2(1+|r_2|^2A_{1,2}^2)(1+|r_0|^2A_{1,0}^2),
\label{res_current_density}
\end{equation}
where we introduced
\begin{equation}
G_{20}=\frac{d_M}{(1-A_{1,2}^2 r_2{\tilde r}_M)(1-A_{1,0}^2 r_0 r_M)
-A_{1,0}^2 A_{1,2}^2 r_0 r_2 d_M {\tilde d}_M}.
\label{propagator}
\end{equation}

The factors in the current density given in
Eq.~(\ref{res_current_density}) can be interpreted in a suggestive way
(see also the more general discussion in Ref.~\cite{JWBS}).  The
factor of two appears since we are considering a two particle process
(transfer of the charge $2e$). The factor $(1-A_{1,0}^2)$ is the
probability to enter the normal region at energy $E_0$ and the factor
$(1-A_{1,2}^2)$ is the probability to exit at energy $E_2$. The
propagator $G_{20}$ is taking us from the injection point up to the
exit point. The factor $(1+|r_2|^2 A_{1,2}^2)$ adds the two possible
ways of exiting the normal region. First, we may go out directly when
we come from below (giving the term $1$). Second, we may Andreev
reflect at $E_2$, go further up and be reflected back (in region III)
and then go out at $E_2$ (giving the term $|r_2|^2 A_{1,2}^2$). The
last factor $(1+|r_0|^2 A_{1,0}^2)$ adds the contributions from
injected electron-like quasiparticles (giving the term $1$) and
hole-like quasiparticles (giving the term $|r_0|^2 A_{1,0}^2$).  The
reason for getting the extra factor $|r_0|^2 A_{1,0}^2$ for the
hole-like quasiparticles is that the injection is downwards [into the
coefficient $c_0$ as seen in Eq.~(\ref{abcd})] meaning that the
particle must be reflected at negative energy, and then be Andreev
reflected at $E_0$ before going up in energy and out from the normal
region at $E_2$.

\begin{figure}[t]
\centerline{\psfig{figure=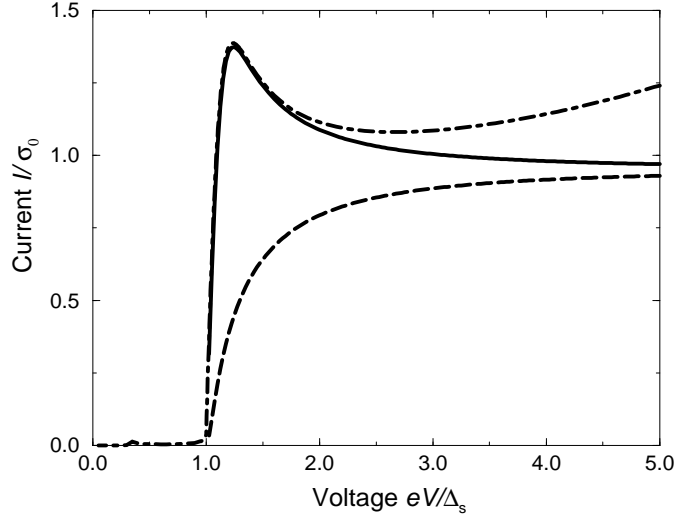,width=10cm}}
\caption{Comparing the current contribution from the two-particle 
  process (solid line) with the complete IV-curve (dot-dashed line),
  we see that the peak is due to this particular process. For
  comparison, we include the current (dashed line) from the bare MGS
  resonance (letting $r_0=r_2=0$). Taking into account the reflections
  $r_0$ and $r_2$ gives rise to boundary resonances, which at the
  voltage $eV=\Delta_s=0.2\Delta_0$ overlap with (and therefore
  broadens) the MGS resonance and produce the current peak.}
\label{IV-curves}
\end{figure}

Eqs.~(\ref{res_current_density}) and (\ref{propagator}) describe the
current density from quasiparticles incoming from the left electrode
at energy $E\approx -eV$, resonantly transmitted through MGS, leaving
the normal region into the left electrode at energy $E_{2} \approx
eV$.  This process is not possible for voltages below $eV=\Delta_{s}$,
because then either $E$ or $E_{2}$ is within the gap of the left
superconductor. For voltages well above $eV=\Delta_{s}$ we may
approximate the propagator $G_{20}$ by the bare transmission amplitude
$d_{M}$, since $A_{1,0} \approx A_{1,2} \approx 0$.  For the
particular orientation $\alpha=\pi/4$, $|d_M|^2$ is equal to
$1/(1+E^2/\Gamma^2)$, i.e. a Breit-Wigner resonance of width
$\Gamma={\cal D}|\Delta(\theta)|/[2(1-{\cal D})]$.  The total
contribution to the current, after integration over energy, is then
$I\propto \Gamma$, i.e. $I$ is proportional to ${\cal D}$, which is of
the same order of magnitude as the single particle current.  This
means that the main current (current proportional to ${\cal D}$) has
an onset at $eV=\Delta_{s}$, due to the resonant two-particle process.
Fig.~\ref{IV-curves} confirms this picture: we see that the resonant
two-particle process (solid line) gives the main contribution to the
total current (dot-dashed line) for voltages between $\Delta_{s} < eV
< \Delta_{0}$.  For voltages $eV > 2\Delta_{s}$ the single particle
contribution is noticeable. We note in passing that the above effect
is similar to the effect of a Breit-Wigner resonance in the normal
region of an $s|s$ junction~\cite{JBSW}.

Close to the main current onset at $eV=\Delta_{s}$, the magnitude of
the Andreev reflection amplitudes $|A_{1,0}|$ and $|A_{1,2}|$ are both
close to unity. This together with the reflections from $\pm\infty$ in
energy (given by $r_0$ and $r_2$) give rise to an additional resonance
in $G_{20}$ (the denominator gives rise to a singularity at the gap
edge). This {\it boundary} resonance is quite weak and does not change
the order of magnitude of the current on its own, since it is
connected to the divergency in the superconducting density of states
which is integrable. At the onset ($eV=\Delta_s$) the boundary
resonance overlaps with the MGS resonance. The overlap broadens the
bare transmission amplitude which results in a current peak. The
boundary resonance is due to the reflection from $\pm\infty$ in
energy, so by removing them by hand (putting $r_{0}=r_{2}=0$) we find
that the current peak dissappears and is replaced by a smooth onset
(the dashed line in Fig.~\ref{IV-curves}).

We note that the ${\mbox N}|d_{\pi/4}$ case is easily reached by
letting $\Delta_s\rightarrow 0$. In this limit the current onset will
be at $eV=0$ and the current peak disappears since we lose the
boundary resonances ($r_0$ and $r_2$ $\rightarrow$ $0$ when
$\Delta_s\rightarrow 0$). The onset at $eV=0$ result in a
ZBCP~\cite{Hu}-\cite{Alff1}.

\section{Zero-bias conductance peak in the $d_{\alpha}|d_{-\alpha}$ junction}

In recent experiments on grain boundary junctions of HTS a ZBCP has
been found~\cite{Alff2,Alff3}. Within our model a ZBCP can only be
found in rather transparent $d_{\alpha}|d_{-\alpha}$ junctions as
previously reported~\cite{HLJW}.  Generally, the current contribution
from a process of order $n$ is (without resonances) proportional to
the transparency ${\cal D}$ of the junction to the power of
$n$~\cite{SBW,BSBW,JWBS}.  In the limit of small bias voltage, the
current contributions are from high-order processes, meaning that the
current is extremely small without any resonances. This picture is
changed by the MGS resonance.

\begin{figure}[t]
\centerline{\psfig{figure=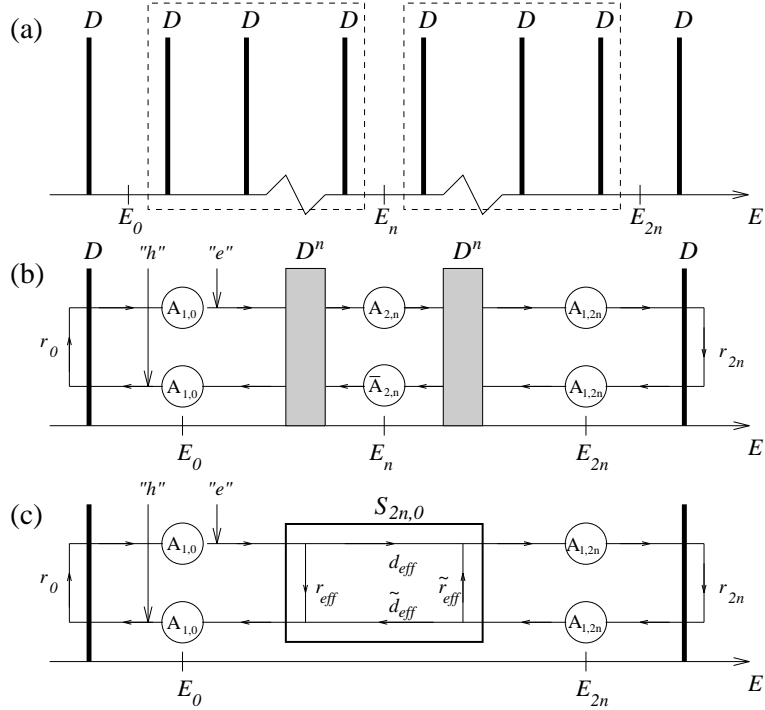,width=10cm}}
  \caption{Mapping of the $2n$ particle process involving MGS
    contributing to the current in the high transparency
    $d_{\alpha}|d_{-\alpha}$ junction at small voltage. The $n$
    barriers on each side of MGS shown in (a) are in (b) collected
    into effective barriers with transparency ${\cal D}^n$. The
    resonant transport through MGS can then be described by an
    effective scattering matrix as drawn in (c). The bare transmission
    resonance (described by $d_{{\mbox eff}}$) is broadened by
    boundary resonances (due to the reflections $r_0$ and $r_{2n}$ at
    the boundaries) producing an enhanced current at small voltage,
    i.e a ZBCP.}
  \label{dd_map_fig}
\end{figure}

In Fig.~\ref{dd_map_fig} we have drawn the map of the scattering state
contributing to the current at small bias voltage. The injected
quasiparticles undergo multiple Andreev reflections and exit the
normal region above the gap. At some particular voltage, the lowest
order process contributing to the current, will contain $2n$ Andreev
reflections and MGS will be reached, after $n$ Andreev reflections, at
the sideband energy $E_n$. The path from the injection point up to MGS
consists of $n$ passings through the barrier. In
Fig.~\ref{dd_map_fig}b we have collected this path into an effective
first-order path with a barrier of height ${\cal D}^n$. The same is
done for the path from MGS up to the exit point. The map in
Fig.~\ref{dd_map_fig}b is analagous to the map in
Fig.~\ref{sd_map_fig}a, meaning that we should expect a Breit-Wigner
resonance of width $\Gamma_{{\mbox eff}}\propto D^n$, broadened by the
boundary resonances. In Fig.~\ref{dd_map_fig}c the middle region
(corresponding to region I in the discussion of the $s|d_{\pi/4}$
junction) is collected into a scattering matrix. We can from this
picture immediately write down a formal expression for the current,
equivalent to Eqs.~(\ref{res_current_density}) and (\ref{propagator}),
in terms of the effective reflection and transmission amplitudes
through MGS ($r_{{\mbox eff}}$, ${\tilde r}_{{\mbox eff}}$, $d_{{\mbox
    eff}}$, and ${\tilde d}_{{\mbox eff}}$), and the reflection
amplitudes from $\pm\infty$ in energy ($r_0$ and $r_{2n}$).  Since the
process under consideration is of order $2n$, the expression for the
current will contain the factor $2n$. Due to the MGS resonance, the
current density will therefore contain peaks of height $2n$ and width
$\Gamma_{{\mbox eff}}$, which after the integration over energy give
rise to an enhanced current at small voltage resulting in a ZBCP.
Since the peak width scales with ${\cal D}^n$ (neglecting broadening
effects due to boundary resonances), the current contribution will be
considerable only in junctions with high transparency~\cite{HLJW}.

\section{Summary}

We have discussed the effects of the midgap state on current transport
through voltage-biased Josephson junctions of $d$-wave superconductors
and demonstrated that in connection with multiple Andreev reflection
the MGS acts as a transmission resonance in energy space.

Depending on the orientation of the $d$-wave superconductors and the
transparency of the junction, resonant transport through MGS
influences the current in different ways. In low transparency
$s|d_{\pi/4}$ junction the MGS resonantly enhances the two-particle
current, which gives the main contribution to the current for voltages
between $\Delta_{s} < eV < \Delta_{0}$.  At the onset an overlap of
boundary resonances, connected to the peaks in the density of states,
and the MGS resonance results in a current peak. In the same way, the
MGS resonantly enhances the two-particle current in $d|d$-junctions
which, depending on the orientations of the superconductors, results
in a current peak at finite voltage~\cite{Hurd}. Since the gap has
nodes in both electrodes there is no clear onset of the resonant
current. In addition, a smooth background contribution from the
single-particle current at all voltages makes the effect less
pronounced compared to the $s|d_{\pi/4}$ junction. The relation
between our results and experiments was discussed in Ref.~\cite{HLJW}.

In high transparency $d_{\alpha}|d_{-\alpha}$ junctions there are
resonant $2n$-particle processes involving the MGS, corresponding to
tunneling through an effective symmetric double barrier structure in
energy space. These symmetric processes give a huge contribution to
the current, resulting in finite current at low voltages, as in
ballistic $s|s$-juntions. This may explain the zero-bias conductance
peaks seen in experiments on grain boundary junctions of HTS
superconductors~\cite{Alff2,Alff3}.

\end{document}